\documentclass[a4paper,11pt]{article}
\usepackage{pos}

\usepackage{slashed,mathtools}
\newcommand{\klammer}[1]{ \left( #1 \right)  }

\newcommand{\nm}{n_-}
\newcommand{\np}{n_+}

\title{Endpoint factorization and next-to-leading power resummation of gluon thrust}
\ShortTitle{Endpoint factorization and NLP resummation of gluon thrust}

\author*[a,b]{Martin Beneke}
\author[a]{Mathias Garny}
\author[c]{Sebastian Jaskiewicz}
\author[a,b]{Julian Strohm}
\author[d]{Robert Szafron}
\author[e,f]{Leonardo Vernazza}
\author[g]{Jian Wang}

\affiliation[a]{Physik Depertment T31, Technische Universit\"at M\"unchen,\\
  James-Franck-Stra{\ss}e 1, D-85748 Garching, Germany}

\affiliation[b]{Excellence Cluster ORIGINS, Technische Universit\"at M\"unchen,\\
D-85748 Garching, Germany}

\affiliation[c]{Institute for Particle Physics Phenomenology, Durham University,\\
  South Road, Durham DH1 3LE, United Kingdom}
  
\affiliation[d]{Department of Physics, Brookhaven National Laboratory,\\
  Upton, NY 11973, U.S.A}

\affiliation[e]{INFN, Sezione di Torino,
  Via P. Giuria 1, I-10125 Torino, Italy}
  
\affiliation[f]{Nikhef, 
Science Park 105, NL-1098 XG Amsterdam, Netherlands}

\affiliation[g]{School of Physics, Shandong University,
Jinan, Shandong 250100, China}

\abstract{Endpoint divergences in the convolution integrals appearing in 
next-to-leading-power factorization theorems prevent a straightforward application
of standard methods to resum large 
logarithmic power-suppressed  corrections in collider physics. 
We study the   power-suppressed configuration of the thrust 
distribution in the two-jet region, where a gluon-initiated jet recoils against 
a quark-antiquark pair. With the aid 
of operatorial endpoint factorization conditions, we derive  
a factorization formula where the individual terms are free from 
endpoint divergences and can be written in 
terms of renormalized hard, (anti) collinear, and soft functions in four 
dimensions. This framework enables us to perform the
first resummation of the 
endpoint-divergent SCET$_{\rm I}$ observables at the leading 
logarithmic accuracy using 
exclusively renormalization-group methods. }

\FullConference{%
  Loops and Legs in Quantum Field Theory - LL2022,\\
  25-30 April, 2022\\
  Ettal, Germany
}

\begin{document}

\renewcommand{\hookAfterAbstract}{%
\par\bigskip\bigskip
\textsc{IPPP/22/50}, 
\textsc{TUM-HEP-1411/22} 
}
\maketitle

\section{Introduction}

\noindent The endpoint factorization and next-to-leading power (NLP)
resummation of gluon thrust employing renormalization group (RG) methods has
recently been presented in~\cite{Beneke:2022obx}. This proceeding highlights the 
key insights alongside showcasing the main results.
For precise definitions and technicalities of the derivation, we direct the interested reader to the original 
publication. 

In the past, hadronic event shape variables
in the two-jet region were of great importance
in the development of diagrammatic resummation methods for QCD~\cite{Catani:1991kz,Catani:1992ua}.
Later, the advent of soft-collinear effective theory (SCET) enabled the 
accuracy of resummation to be further improved. This was first shown
for the thrust variable $T$~\cite{Becher:2008cf}, defined as 
\begin{align}
T=\mbox{max}_{\vec n}\,\frac{\sum_{i}
\left|\vec{p_{i}}\cdot\vec{n}\right|}
{\sum_{i}\left|\vec{p_{i}}\right|},
\end{align}
where the index $i$ sums over all final state hadrons (partons).
As 
$\tau=1-T \to 0$,\, 
back-to-back jets are formed by the particles, and 
large logarithms $\ln\tau$ appear at every order in $\alpha_s$, 
signalling a breakdown in perturbation theory.  This phenomenon has been 
scrutinized in great detail for the quark-antiquark two-jet process
that contributes at leading power (LP) in the $\tau$ expansion. 

\begin{figure}
\begin{center}
\includegraphics[width=0.24\textwidth]{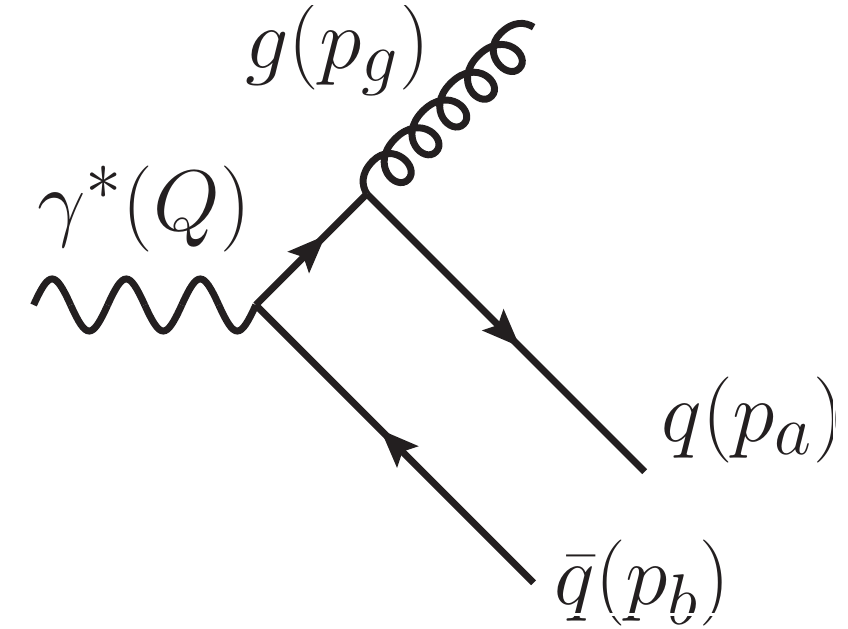}\qquad\qquad
\includegraphics[width=0.24\textwidth]{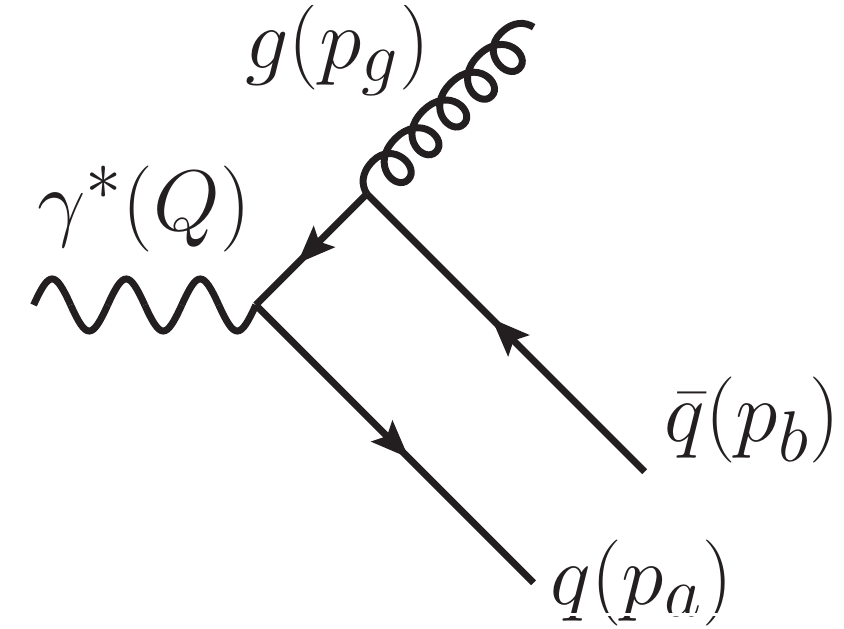}
\caption{\label{fig:QCD}  
Diagrams contributing to gluon thrust
at leading $\alpha_s$ order in QCD.}
\end{center}
\end{figure}

In the work presented here, we instead focus on the ``gluon thrust'' phase-space region
where at leading order (LO) the gluon recoils a quark-antiquark pair
\begin{equation}
e^+ e^-\to\gamma^*\to [g]_{c} + [q\bar q]_{\bar c}\,.
\end{equation}
The gluon and the quark-antiquark jets are chosen to be in the
collinear and anticollinear directions, respectively. As shown in figure~\ref{fig:QCD}, 
this process begins at $\mathcal{O}(\alpha_s)$
and it is of particular interest in our investigations as in the limit of $\tau\to 0$ the LP 
$[\ln\tau/\tau]_+$ soft-gluon behaviour is absent. 
Instead, the process begins at NLP in the $\tau$ expansion, 
with the leading term being $\alpha_s\ln \tau$.

The study of power corrections in a multitude of contexts has 
recently gathered considerable attention.
However, as we discuss below, further rapid progress has thus far been hindered by the 
ubiquitous appearance of endpoint divergences in convolution integrals between the 
hard, (anti-) collinear, and soft functions in the NLP factorization theorems, such as for the Drell-Yan threshold~\cite{Beneke:2019oqx}.
We address this key conceptual issue in the context of gluon thrust, as it allows us to focus on the problem of endpoint divergences without needing to  address additional difficulties such as factorization of parton distribution functions, which would affect Drell-Yan and DIS processes. 

The result for the all-order logarithmic structure of
``gluon thrust'' at the double logarithmic (DL) accuracy 
was first written down in~\cite{Moult:2019uhz} and 
later derived from $d$-dimensional consistency 
relations in~\cite{Beneke:2020ibj}. 
An interesting feature of this result is the unconventional ``quark'' 
Sudakov form factor, which arises due to the colour mismatch 
between the back-to-back energetic particles when 
either the quark or anti-quark in the 
$q\bar q$-jet becomes soft, causing the double logarithms
to be proportional to the difference of the colour charges,
$C_A-C_F$. Despite the progress in the description of this 
process within SCET \cite{Moult:2019uhz,Moult:2019mog,Beneke:2020ibj},
it has not yet been possible to improve the resummation 
accuracy beyond DL order due to the presence of endpoint-divergences
in the relevant convolution integrals.

In this work, we develop the refactorization ideas employed 
for the DIS process \cite{Beneke:2020ibj}, and Higgs decay to two photons 
through light-quark loops \cite{Liu:2019oav,Liu:2020wbn} in order to 
derive a SCET$_{{\rm{I}}}$ NLP factorization theorem valid in $d=4$ dimensions. To arrive
at this result, we make use of standard factorization within SCET combined
with endpoint factorization.

\section{Heuristic  discussion}
\label{sec:heuristic}

\noindent In this section, we present a sketch of the
factorization formula in the EFT framework, which aids in
motivating the endpoint rearrangement and subtraction 
terms introduced below. At $\mathcal{O}(\alpha_s)$, 
there are two ways to induce the gluon jet:  
\begin{enumerate}
\item[I] Both the quark and the anti-quark can carry large anti-collinear 
momentum and create a single jet recoiling against the collinear gluon.
\item[II] Either the quark or the anti-quark is anti-collinear and balances
the collinear gluon momentum, in which case the other fermion is soft.
\end{enumerate}
Both 
situations provide identical
power suppression in $\tau$, and the
further evolution of the process is governed by LP interactions. However, the 
separation into possibilities~I~and~II introduces 
an ambiguity in what is precisely meant by the ``soft'' and ``anti-collinear'' modes. 
If we begin with configuration~I~and lower the large anti-collinear momentum 
of either the quark or the anti-quark, eventually it will become soft, at which
point it should in fact be counted as a contribution to situation~II rather than~I.
As long as the definition of the jet is infrared safe, this separation is not an
issue for fixed-order computations. It is a problem, however, if the goal is to 
perform resummation since, in this case, a clear separation of soft and (anti) collinear 
modes is required in order to disentangle large logarithms into single-scale
pieces. This tension between the mathematical description of the modes and the physical 
picture is the origin of the endpoint-divergences which appear in convolution integrals
of the NLP factorization theorems. 

\begin{figure}
\begin{center}
\includegraphics[width=0.3\textwidth]{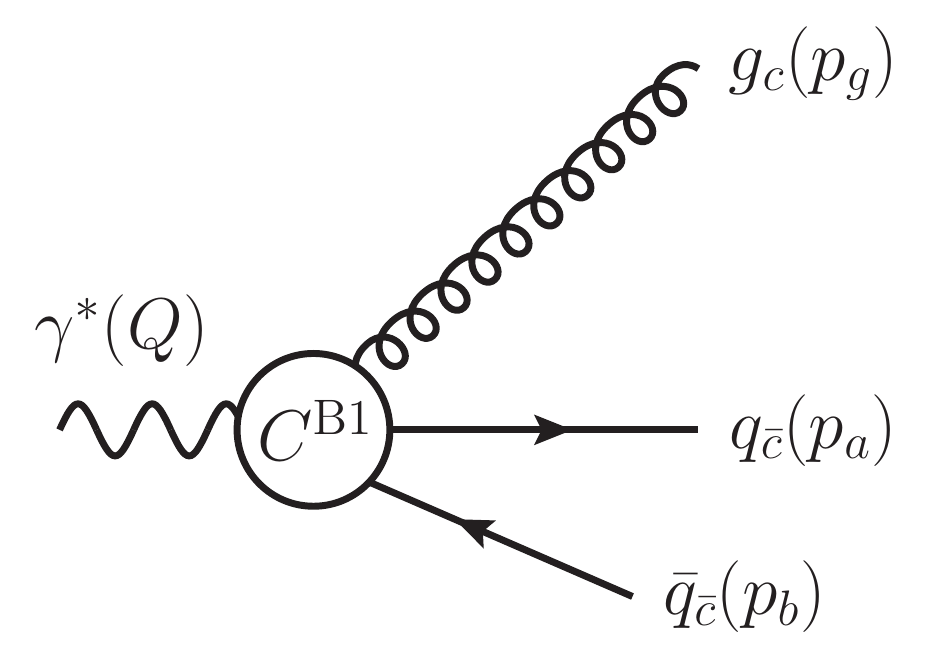}\hspace{0.15cm}
\includegraphics[width=0.3\textwidth]{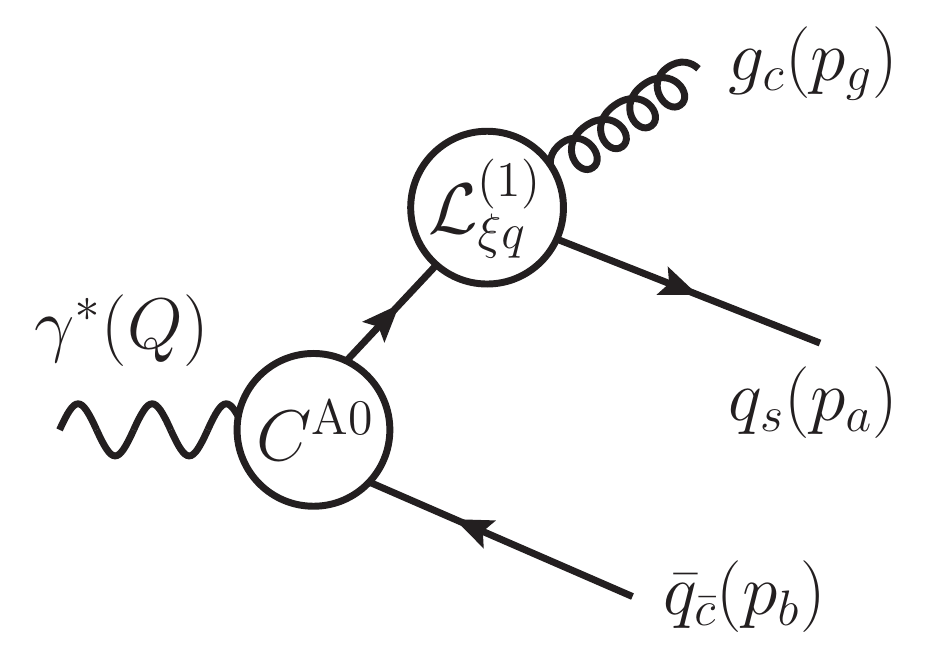}\hspace{0.15cm}
\includegraphics[width=0.3\textwidth]{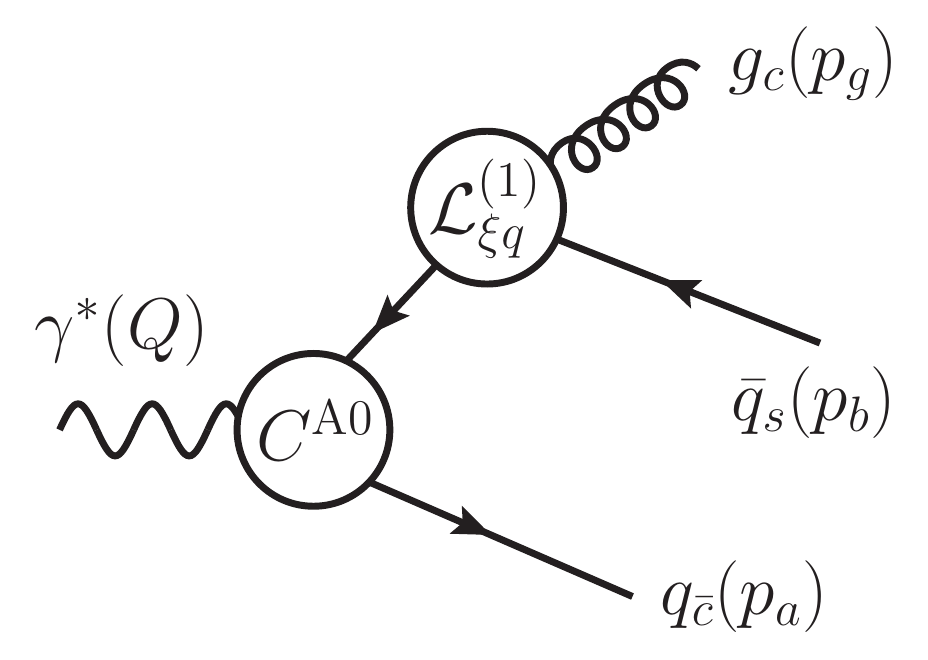}
\caption{\label{fig:SCET} Representation of the 
gluon-thrust amplitude in the two-jet region in SCET. 
}
\end{center}
\end{figure}

We now focus on the Feynman diagrams in figure~\ref{fig:QCD} and
analyze possibilities I and II from the point of view of SCET. 
Starting with I, later referred to as ``B-type'', we see that the 
internal propagator in the left diagram carries hard momentum 
$q_a=p_g + p_a$ (since $p_g$ is collinear and  $p_a$ anti-collinear). 
As the intermediate propagator is hard,  $q_a^2\sim Q^2$, it is integrated out,
and the hard scattering vertex in SCET directly produces a $q\bar{q} g$ state.
This is shown in the left-most diagram of figure~\ref{fig:SCET}.  
Only the total momentum of the $q\bar q$ anti-collinear pair is fixed by momentum
conservation, while the amplitude depends on the fraction of the momentum carried by 
each particle. In the case where one of these fractions becomes vanishingly 
small, the corresponding parton is effectively soft and should be counted
as possibility II, which we later refer to as the ``A-type'' contribution. Here, the 
$q_a$-intermediate propagator, or  $q_b$ for the soft anti-quark case, 
ceases to be hard. Therefore, the hard scattering vertex is the LP
$\gamma^*\to q\bar {q}$ process and the entire momentum of the energetic
quark (or anti-quark) is subsequently transferred to the gluon, rendering the
daughter fermion soft. In SCET, such a situation is described by an insertion of a
power-suppressed SCET Lagrangian interaction term $\mathcal{L}^{(1)}_{\xi q}$,
which constitutes soft (anti-) quark emission, as depicted in the middle 
and right-most diagrams of figure~\ref{fig:SCET}. 

The fact that there exists an overlap region where the limits of two different
expressions can describe the same physical process is central to the idea of 
endpoint factorization. 
Before we make these concepts concrete, let us start with the following schematic 
discussion. In Laplace space, the factorization theorem for two-hemisphere invariant mass
distribution of gluon thrust takes the following form: 
\begin{align}
\frac{1}{\sigma_{0}}\frac{\widetilde{d\sigma}}{ds_R ds_L}
&=\int_{0}^{\Lambda} d\omega d\omega'\,
\left|C^{A0}\right |^{2}\times
\mathcal{J}^{(\bar{q})}_{\bar{c}}\times 
\mathcal{J}_{c}\left(\omega,\omega'\right)
\otimes S_{\rm NLP} \left(\omega,\omega'\right)
\nonumber \\&+
\int_{\Lambda/Q}^{1-\Lambda/Q} drdr'\,
C^{B1}(r) C^{B1}(r')^{*}\otimes
\mathcal{J}^{q\bar q}_{\bar{c}}\left(r,r'\right)\times 
\mathcal{J}_{c}^{(g)}\times S^{(g)}\,.
\label{eq:ffschematic}
\end{align}
In this formula, the hard matching coefficients are denoted by $C$, the
jet functions by $\mathcal{J}$ and the soft functions by $S$.
In the arguments of functions, we only retain the dependence on convolution variables
which contain divergent integrals. The $\omega$ and $\omega'$ soft-collinear convolutions
diverge logarithmically for $\omega,\omega'\to \infty$, and the $r$, $r'$
hard-anti-collinear convolution integrals are logarithmically divergent for
$r,r'\to 0,1$. 

The soft function $S_{\rm NLP}$ contains the soft quark from 
situation II. In the overlap region, this soft quark carries a 
{\em large} soft momentum $\omega$ and could be thought of as being
a part of the $\mathcal{J}^{q\bar q}_{\bar{c}}$  anti-collinear function 
appearing in the bottom line of~\eqref{eq:ffschematic}, with a {\em small} 
anti-collinear momentum fraction $r$. Taking away this quark from 
$S_{\rm NLP}$ leaves behind only $S^{(g)}$, therefore in total $S_{\rm NLP}\to S^{(g)}$,
$\mathcal{J}^{(\bar{q})}_{\bar{c}} \to \mathcal{J}^{q\bar q}_{\bar{c}}$, and the 
hard process changes from A0-type to B1-type. 
Hence, in these limits, the  {\em integrands} of the two terms in \eqref{eq:ffschematic} should be identical. 
This fact allows us, in the singular limits, to perform a rearrangement at the {\em integrand} level, such that both terms are individually finite.
At this stage, we can 
employ 
standard RG techniques to resum the logarithms in the hard, jet, and soft functions, as we show in more detail in the following sections. 

\section{Bare factorization theorem}

\noindent Before focusing on endpoint factorization, we state 
the results of the derivation of the $d$-dimensional SCET factorization formula for gluon thrust \cite{Beneke:2022obx}.
To start, we integrate out the hard modes and match the 
electromagnetic current to 
\begin{eqnarray}
\bar\psi\gamma_\perp^\mu\psi(0) &=& \int dtd\bar{t}\,
\widetilde{C}^{\rm A0}(t,\bar{t})\,
\bar{\chi}_c(t\np)\gamma_\perp^\mu\chi_{\bar c}(\bar t\nm) 
+ (c\leftrightarrow \bar{c})
\nonumber\\
&&+\,\sum_{i=1,2}\int dt d\bar{t}_1d\bar{t}_2\,
\widetilde{C}^{\rm B1}_i(t,\bar{t}_1,\bar{t}_2)\,
\bar{\chi}_{\bar c}(\bar{t}_1\nm)\Gamma_i^{\mu\nu}\mathcal{A}_{c\perp\nu}
(t\np)\chi_{\bar c}(\bar{t}_2\nm)
+\ldots \qquad
\label{eq:hardmatching}
\end{eqnarray}
In the first line, we have the LP hard scattering 
vertex, which produces a back-to-back quark-antiquark pair. It
contributes to gluon thrust through a time-ordered 
product with the  $\mathcal{O}(\lambda)$ suppressed SCET 
Lagrangian  \cite{Beneke:2002ph} term
\begin{align}
\mathcal{L}_{\xi q}(x)=\bar{q}_s(x_{-})\slashed{\mathcal{A}}_{c\perp}(x)\chi_{c}(x)+\rm{h.c.}\,,
\label{eq:Lxiq}
\end{align} 
which transforms a collinear quark into a collinear gluon and 
a soft quark. The soft quark argument $x_\mp^\mu$ is defined as  $x_\mp^\mu = (n_\pm\cdot x)\frac{n_\mp^\mu}{2}$. In the second line of \eqref{eq:hardmatching},
we have the $\mathcal{O}(\lambda)$ suppressed 
``B-type'' SCET operator, which produces
a collinear gluon and an 
anti-collinear quark-antiquark pair directly, see the left-most diagram in figure~\ref{fig:SCET}.
The relevant Dirac structures in the B-type operator are 
\begin{equation}
\Gamma_1^{\mu\nu}=\frac{\slashed n_{-}}{2}\gamma_{\perp}^{\nu}\gamma_{\perp}^{\mu}\,,
\qquad\quad
\Gamma_2^{\mu\nu}=\frac{\slashed n_{-}}{2}\gamma_{\perp}^{\mu}\gamma_{\perp}^{\nu}\,.
\label{eq:BtypeDirac}
\end{equation}
The remaining steps of the derivation are fairly standard. 
The collinear, anti-collinear and soft fields are decoupled at LP by the 
redefinition of the collinear field with the 
soft Wilson line \cite{Bauer:2001yt}
$Y_{n_-}(x)=\mathcal{P} \exp\left[i g_s \int_0^\infty ds n_-A_s(x+sn_-) \right]$. 
We then square the matrix elements, 
sum and integrate over all possible final states.
Since the final 
state  $|X\rangle = |X_c\rangle|X_{\bar c}\rangle 
|X_s\rangle$ is made up of corresponding modes, the matrix element factorizes into 
collinear, anti-collinear and soft functions. 
Following this prescription for the two terms in \eqref{eq:hardmatching} gives rise 
to the A-type and B-type contributions to the factorization formula, for which we present
the results separately.   We give results for the two-hemisphere 
invariant-mass distribution, which is related to the thrust distribution by
 \begin{equation}
 \frac{d\sigma}{d\tau} =
\int dM_R^2dM_L^2 \,\delta\!\left(\tau-\frac{M_R^2+M_L^2}{Q^2}
\right)\frac{d\sigma}{dM_R^2 dM_L^2}\,. 
 \label{eq:taudist}
 \end{equation}
For the A-type contribution we find
\begin{eqnarray}
\frac{1}{\sigma_0}\frac{d\sigma}{dM_R^2 dM_L^2}|_{\rm A-type} &=& 
\frac{2 C_F}{Q}\,f(\epsilon)\,
|C^{\rm A0}(Q^2)|^2 \int_0^\infty dl_+ dl_-\,
\int d\omega d\omega'\,
\mathcal{J}_{\bar c}^{(\bar q)}(M_R^2-Q l_+)
\nonumber\\
&&\hspace*{-4cm}\times\,
\bigg\{\,\mathcal{J}_{c}(M_{L}^{2}-Ql_-,\omega,\omega')
\,S_{\rm NLP}(l_+,l_-,\omega,\omega')
+\,\widehat{\mathcal{J}}_{c}(M_{L}^{2}-Ql_-,\omega,\omega')\,
\widehat{S}_{\rm NLP}(l_+,l_-,\omega,\omega')\,\bigg\}\,,
\label{eq:Atypefactformula}
\end{eqnarray}
and for the B-type
\begin{eqnarray}
\frac{1}{\sigma_0}\frac{d\sigma}{dM_R^2 dM_L^2}|_{\rm B-type} &=& 
\frac{2 C_F}{Q^2} \,f(\epsilon)\int_0^\infty dl_+ dl_-\,
\sum_{i,i'=1,2} \int dr dr'\,
C^{\rm B1*}_{i'}(Q^2,r')C^{\rm B1}_{i}(Q^2,r) 
\\\nonumber
&&\hspace*{-4cm}\times 
\bigg\{ 
\delta_{i i'} \mathcal{J}_{\bar{c}}^{q\bar{q}(8)}(M_R^2-Ql_+,r,r^\prime)
+ (1-\delta_{i i'}) \mathcal{\widehat{J}}_{\bar{c}}^{q\bar{q}(8)}(M_R^2-Ql_+,r,r^\prime)
\bigg\} 
\mathcal{J}_{c}^{(g)}(M_L^2-Q l_-)\,S^{(g)}(l_+,l_-)\,. 
\label{eq:Btypefactformula}
\end{eqnarray}
In the above equations, $f(\epsilon)$ is a $d$-dimensional factor with 
$f(0)=1$ in four dimensions. The precise operator definitions for
all the functions, along with the lowest-order results, can be found
in sections 3.1 and 3.2 of \cite{Beneke:2022obx} for the A and B-type
parts, respectively.

\subsection{Tree-level evaluation}
\label{sec:treelevelevaulation}

\noindent It is instructive to investigate the structure of the expressions already
at the lowest order. We take the tree-level results for all required
functions from \cite{Beneke:2022obx} and insert them into 
equations \eqref{eq:Atypefactformula} and
\eqref{eq:Btypefactformula} which yields the following expressions.  The
A-type soft-quark term is 
\begin{eqnarray} 
\label{eq:AtypesqTree}
\frac{1}{\sigma_0}\frac{d^2\sigma}{dM^2_R\,dM^2_L}|_{
\scriptsize{ \begin{array}{l}
$\rm A--type$,\\[-0.1cm] \rm tree \end{array}}} 
\!=\frac{\alpha_s}{4\pi} 
\frac{2C_F}{Q^2}  \frac{f(\epsilon)e^{ \epsilon\gamma_E}}{\Gamma(1-\epsilon)}
\bigg[ \frac{1}{\epsilon}\delta^+(M^2_L)
\left(\frac{({M^2_R})^2}{{Q}^2\mu^2}\right)^{\!-\epsilon}\!
-   \frac{\delta^+(M^2_R )}{1-\epsilon}
\left(\frac{ (M^2_L)^2}{Q^2\mu^2}\right)^{\!-\epsilon}\bigg], 
\end{eqnarray}
where the $\epsilon$-pole originates from the logarithmic divergence
in the $\int_{l_+}^\infty d\omega/\omega$ integral for large  $\omega$
values. The  soft-antiquark term is identical to the above. 
Turning to the B-type term, we find
\begin{eqnarray}
\label{eq:BtypeTree}
\frac{1}{\sigma_0}\frac{d^2\sigma }{dM^2_L\,dM^2_R}
|_{\scriptsize{  \begin{array}{l}
$\rm B--type$,\\[-0.1cm] \rm tree \end{array}}}
\!= \frac{\alpha_s}{4\pi}\,
\frac{4C_F}{Q^2 }\,f(\epsilon)\,\bigg\{
-\frac{1}{\epsilon} + 
\frac{\epsilon}{(1-\epsilon)^2 }\,\bigg\} \, \delta^+(M^2_L)
 \bigg(\frac{M^2_R}{\mu^2e^{ \gamma_E}} \bigg)^{\!-\epsilon}
\frac{\Gamma (2-\epsilon)}{\Gamma (2-2 \epsilon)}\,.
\end{eqnarray}
Here, the  $\epsilon$-pole is due to a logarithmic divergence
in the integral over the momentum fraction $r$ as $r\to 0$ and 
$r\to 1$. Summing both contributions, and taking the limit $\epsilon \to 0$,
we arrive at
\begin{eqnarray}
\label{eq:treeresultA+B}
\frac{1}{\sigma_0}\frac{d^2\sigma }{dM^2_L\,dM^2_R} &=&
\frac{\alpha_s C_F}{\pi}\,\frac{1}{Q^2 }
\,\bigg\{\delta^+(M^2_L) \left[\ln\frac{Q^2}{M_R^2} -1\right] 
- \delta^+(M^2_R)\bigg\}\,.
\end{eqnarray}
After accounting for the soft-antiquark contribution 
to the A-type term, the $\epsilon$-poles cancel in 
the sum of A-type and B-type expressions, as it is needed since 
gluon-thrust is an infrared safe observable. We also reproduce 
the coefficient $-\alpha_s C_F/\pi$ of the $\ln \tau$ term in~\cite{Moult:2016fqy}
after converting to thrust. 

The single logarithm in~\eqref{eq:treeresultA+B} comes from
dimensionally regulated convolution
integrals that are divergent in $d=4$. However, in order to 
perform resummation, we must define renormalized
hard, (anti) collinear, and soft functions and set 
$\epsilon \to 0$ before the convolution integrals are performed. 
This order of proceeding is at this point ill-defined, and in what
follows, we explain how this problem can be solved. Before proceeding, 
we make a helpful tree-level observation. 
Considering~\eqref{eq:Atypefactformula} with tree-level expressions
for the appearing functions, we can expand the integrand in the limit of 
large $\omega, \omega'$ before performing the convolutions and arrive 
at the expression 
\begin{eqnarray} 
\label{eq:AtypesqTreeAsy}
\frac{\alpha_s}{4\pi}\,\frac{2 C_F}{Q^2}\,\delta^+(M^2_L)\, 
\frac{\delta(\omega-\omega')}{\omega\omega'}\,
\frac{f(\epsilon)}{\Gamma(1-\epsilon)}
\,
\omega \left(\frac{M_R^2\omega}{Q\mu^2e^{\gamma_E}}\right)^{\!-\epsilon}
\end{eqnarray}
for the integrand of the $\omega, \omega'$ integral. 
Similarly for the B-type term, inserting tree-level results
into \eqref{eq:Btypefactformula}, expanding the integrand
in the small-$r$ limit, and only then performing the 
$dl_+dl_-$ integrals yields 
\begin{eqnarray}\label{eq:BtypeTreeAsy}
\frac{\alpha_s}{4\pi}
\,\frac{2 C_F}{Q^2 }\,\delta^+(M^2_L)\,
\frac{\delta(r -r' )}{r r'} 
\frac{ f(\epsilon)}{\Gamma(1-\epsilon)}\,r\,
\bigg(\frac{M^2_Rr }{\mu^2e^{\gamma_E}} 
\bigg)^{\!-\epsilon}
\end{eqnarray}
for the $r,r^\prime$ integrand. 
The expression in~\eqref{eq:BtypeTreeAsy} is identical
to one in~\eqref{eq:AtypesqTreeAsy} if we identify 
$r=\omega/Q$, $r'=\omega'/Q$.
From the heuristic discussion given in section~\ref{sec:heuristic},
it is apparent that the agreement is not a coincidence, but rather 
it must hold to all orders in $\alpha_s$ expansion. We formalize
these statements next.

\section{Endpoint factorization}

\noindent In this section, we  use  the coincidence of integrands
of the A-type and B-type terms 
in the asymptotic limits discussed above to 
rearrange and factorize the endpoint contributions so as to 
render the convolution integrals finite. 
Similar techniques were previously used 
in amplitude-factorization problems to derive endpoint factorization
for exclusive $B$ decays to P-wave 
charmonia \cite{Beneke:2008pi}, which uses both SCET and 
non-relativistic QCD, and Higgs decay to two photons 
\cite{Liu:2020wbn}, which is a SCET$_{\rm II}$ process.
Gluon thrust, on the other hand, is a
cross-section level SCET$_{\rm I}$ problem, but 
the main mechanism that achieves endpoint factorization resembles the 
above-mentioned cases.  

\subsection{B1 matching 
coefficients in the soft-collinear limit}

\noindent A key ingredient in the endpoint factorization discussion is the
factorization property of the $C^{{\rm{B}}1}_i$ coefficient function of
the $q\bar q g$ SCET B1 operators in the $r\to 0, 1$ limits. 
The endpoint divergence in this contribution arises because
the intermediate quark or anti-quark goes on-shell, see 
figure~\ref{fig:QCD}. The matching coefficient in the original 
definition is a single-scale, hard function. However, for 
$r\to 0$ (soft quark) and $r\to 1$ (soft anti-quark) cases it 
becomes a two-scale object and can be itself factorized as follows 
\cite{Beneke:2020ibj}:
\begin{align}
& \hskip0.5cm\displaystyle
C_1^{\rm B1}(Q^2,r) = C^{\rm A0}(Q^2)\times 
\frac{D^{\rm B1}(r Q^2)}{r} 
+\mathcal{O}(r^0)\,. 
\label{eq:B1fact1}
\end{align}
Now, the two scales $Q^2$ and $r Q^2$ are separated 
into the LP hard matching coefficient, and a new coefficient
$D^{\rm B1}(r Q^2)$, 
which depends on the endpoint-scale $\sqrt{r} Q$. 
\begin{figure}
\begin{center}
\includegraphics[width=0.65\textwidth]{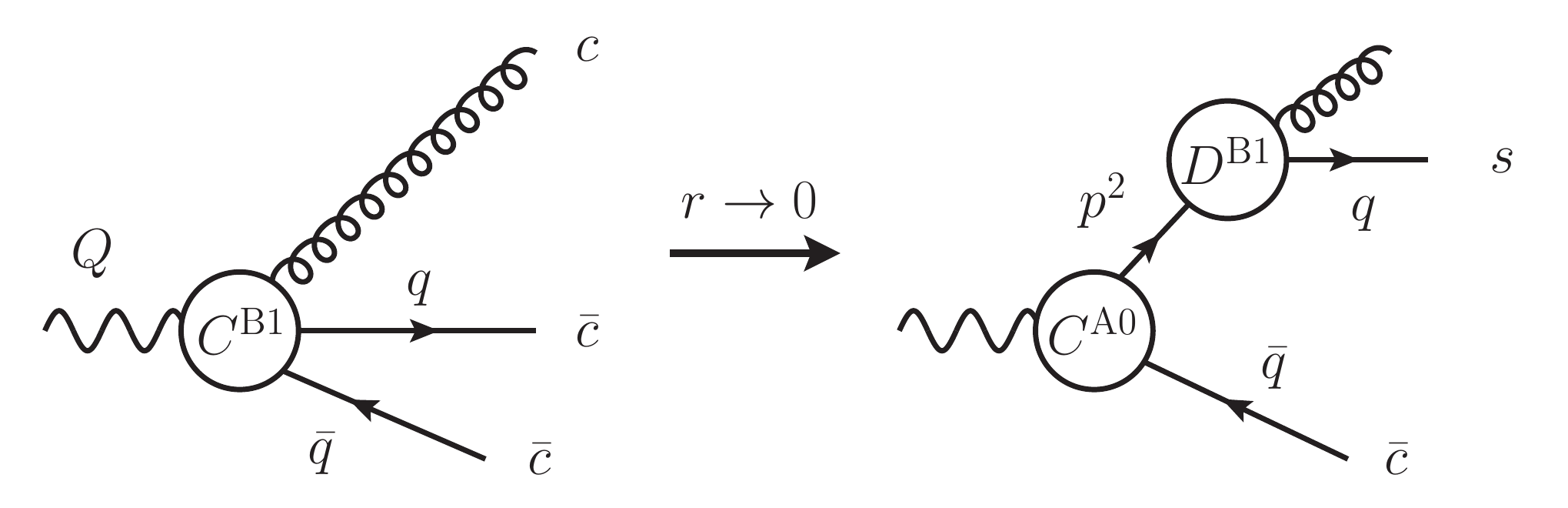}
\caption{
Representation of the 
B1 matching coefficient
factorization  in the soft-quark limit given in~\eqref{eq:B1fact1}.
\label{fig:Cb1refact}}
\end{center}
\end{figure}
The matching coefficient 
$D^{\rm B1}(p^2)$ is a universal function that will appear
in different processes involving soft quark emission.
The DL resummation of $D^{\rm B1}(p^2)$ has been derived  
in \cite{Beneke:2020ibj}. It comprises the all-order 
$(C_A-C_F)^n$ colour coefficient that appears to be 
characteristic in soft quark 
emissions \cite{Vogt:2010cv,Moult:2019uhz,Beneke:2020ibj}. 
The identical coefficient also enters the $ggH$ amplitude with a bottom-quark loop and 
was computed recently at two-loops \cite{Liu:2021mac}. The 
$D^{\rm B1}(p^2)$ coefficient and its evolution equation
can be obtained from the corresponding B1 operator coefficients and anomalous dimension \cite{Beneke:2017ztn,Beneke:2018rbh}
by taking the limit $r\to 0$.  The extraction 
of the anomalous dimension resembles to the derivation 
of the asymptotic kernel for the QED light-meson light-cone 
distribution amplitude \cite{Beneke:2021pkl}, see 
also \cite{Liu:2020wbn}. For the derivation and results, we refer to appendix A of \cite{Beneke:2022obx}.
When the gluon is replaced by a photon, the abelian version 
of $D^{\rm B1}(p^2)$  corresponds to the
jet function in the LP factorization of   $B\to\gamma\ell\nu$ 
\cite{Bosch:2003fc}, and in the NLP endpoint factorization of the 
$H\to\gamma\gamma$ amplitude \cite{Liu:2020wbn}. In the abelian 
case, the evolution up to two loops was inferred from 
renormalization-group consistency of the $B\to\gamma\ell\nu$ 
observable \cite{Liu:2020ydl}. The one-loop evolution kernel
of the jet function has also been obtained by direct 
computation in~\cite{Bodwin:2021epw}. 

\subsection{Endpoint factorization consistency conditions}

\noindent As motivated in section~\ref{sec:heuristic} and confirmed explicitly above 
at $\mathcal{O}(\alpha_s)$, we expect the integrands of the 
A- and B-type terms to have 
the same asymptotic limits to all orders, which is a 
prerequisite for endpoint factorization. Concretely, 
the limit of the anti-collinear momentum $\nm\cdot p_{\bar c} = r Q, r'Q \to 0$ in the B-type term  
should match the limit $\nm\cdot k = \omega,\omega' 
\to \infty$ of the corresponding soft momentum component in 
the A-type term. The above picture is formalized by the two 
refactorization conditions~\cite{Beneke:2022obx}:
\begin{eqnarray}
\mbox{(I)} &\hskip0.5cm 
\mathcal{J}_{c}\left(p^2,\omega,\omega'\right) = 
\mathcal{J}_{c}^{(g)}(p^2) \,
\displaystyle \frac{D^{\rm B1}(\omega Q)}{\omega}
  \frac{{D^{\rm B1}}^*(\omega^\prime Q)}{\omega^\prime} 
+\mathcal{O}\!\left(\frac{1}{\omega^{(\prime)}}\right)\,,
\label{eq:EPfactI}
\end{eqnarray}
and
\begin{eqnarray}
\mbox{(II)} & \hskip-1.5cm 
Q\,\widetilde{\mathcal{J}}^{(\bar{q})}_{\bar{c}}(s_R)\,
\widetilde{S}_{\rm NLP} \left(s_R,s_L,\omega,\omega'\right)\Big|_{\omega^{(\prime)} \to\infty}
\nonumber\\
& \hskip2.5cm \displaystyle
= 
\widetilde{\mathcal{J}}^{q\bar q(8)}_{\bar{c}}\left(s_R,r,r'\right)
\widetilde{S}^{(g)}(s_R,s_L)\Big|_{r^{(\prime)}=\omega^{(\prime)}/Q\to 0}\,.
\end{eqnarray}
For large $\omega$, the soft 
quark field in $S_{\rm NLP}$ turns anti-collinear, 
$\bar{q}_s Y_{n_-} \to \bar{\chi}_{\bar c}$, so it moves 
from $S_{\rm NLP}$ to 
$\mathcal{J}^{q\bar q(8)}_{\bar{c}}$. Removing  $\bar{q}_s Y_{n_-}$ 
from  $S_{\rm NLP}$ leaves the LP soft function 
$S^{(g)}$, and adding it as $\bar{\chi}_{\bar c}$ to 
$\mathcal{J}^{(\bar{q})}_{\bar{c}}$ changes the anti-quark 
jet function into $\mathcal{J}^{q\bar q(8)}_{\bar{c}}$. In total, 
this results in $S_{\rm NLP} \mathcal{J}^{(\bar{q})}_{\bar{c}} \to 
S^{(g)}\mathcal{J}^{q\bar q(8)}_{\bar{c}}$, which is relation 
(II). At the same time, the quark fields in the A-type 
collinear function become highly 
off-shell, which removes them from $\mathcal{J}_{c}(p^2,\omega,\omega^\prime)$. This leaves 
behind only the collinear gluon, so $C^{\rm A0}$ turns 
into $C^{\rm B1}$. Therefore, $|C^{\rm A0}|^2 \mathcal{J}_{c} \to 
|C^{\rm B1}|^2 \mathcal{J}_{c}^{(g)}$, which is relation (I).

\subsection{Endpoint factorization formula}
\label{sec:endpointfact}

\begin{figure}
\begin{center}
\includegraphics[width=0.4\textwidth]{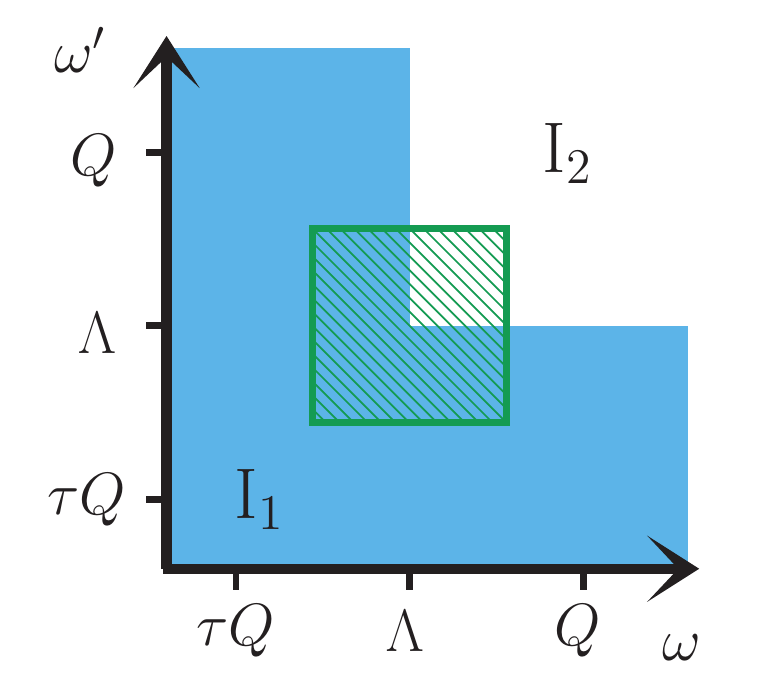}
\includegraphics[width=0.52\textwidth]{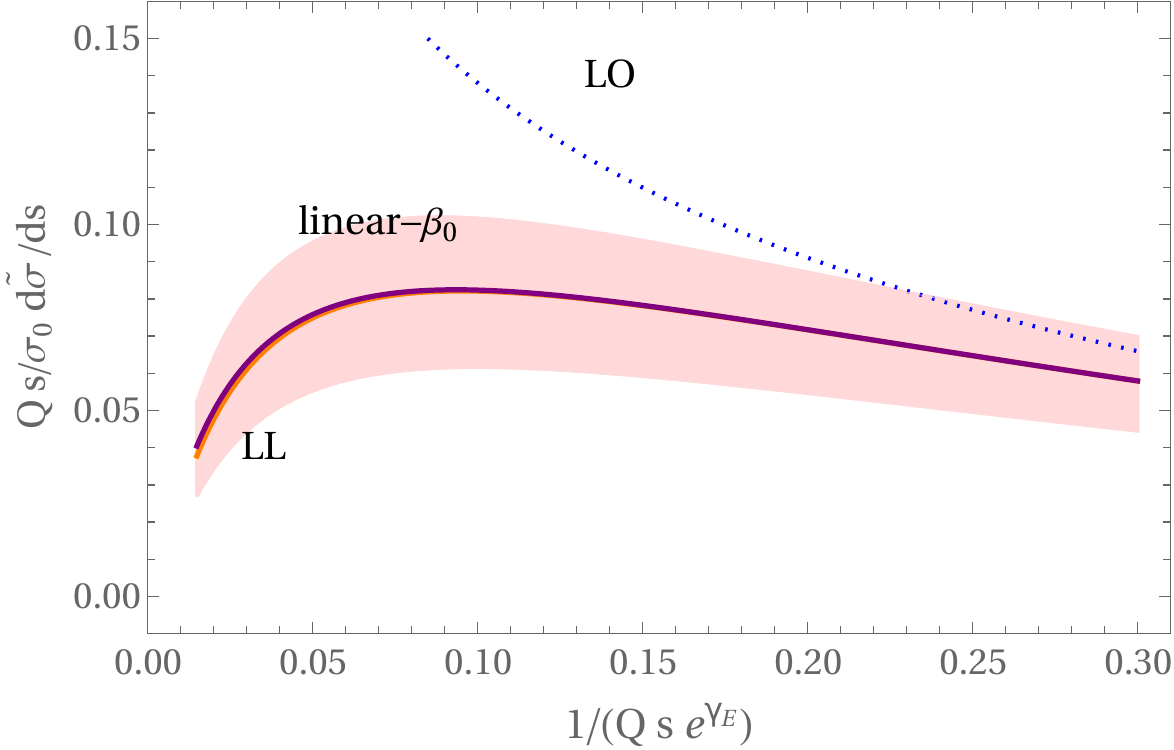}
\caption{\label{fig:overlap}
On the left we show the split of 
\eqref{eq:scaleless_integral} into $I_1+I_2$ according to 
the correspondingly indicated regions in the 
$\omega-\omega'$ plane as described in the text 
below \eqref{eq:scaleless_integral}. In the overlap 
region in green the asymptotic behaviour of the A- 
and B-type term must agree. On the right, we display the 
Laplace-space LL gluon thrust distribution.  Variation of the 
initial scales as described in the text yields the light-red band.
}
\end{center}
\end{figure}

\noindent We can now state the endpoint-finite 
factorization formula. To do so concisely, we use 
the double-bracket notation introduced in \cite{Liu:2019oav} 
to denote the asymptotic behaviours of the various functions. 
In functions of 
$\omega, \omega^\prime$, rescale $\omega \to \kappa 
\omega$, $\omega^\prime \to \kappa 
\omega^\prime$ and take $\kappa\to \infty$. Then 
\begin{eqnarray}
&& \llbracket 
S_{\rm NLP} \left(l_+,l_-,\omega,\omega'\right)
\rrbracket \equiv S_{\rm NLP} \left(l_+,l_-,\omega,\omega'\right)|_{
\mathcal{O}(\kappa^0)}, \label{eq:Sasym}\\
&&\llbracket \mathcal{J}_{c}(p^2,\omega,\omega^\prime)
\rrbracket \equiv \mathcal{J}_{c}(p^2,\omega,\omega^\prime)|_{
\mathcal{O}(\kappa^{-2})}\,.
\end{eqnarray} 
The right-hand 
side of the above equation equals the right-hand side of 
the consistency relation (I).  
Similarly, in functions of 
$r, r^\prime$, rescale $r \to r \kappa$, $r^\prime \to 
r^\prime \kappa$ and take $\kappa\to 0$, {\em or} the 
corresponding rescaling is applied to $\bar r$, $\bar{r}^\prime$. 
Which of the two is meant, will be indicated by the subscript 
0 or 1 on the double bracket. Then 
$\llbracket C^{\rm B1}_1(Q^2,r) \rrbracket_0 \equiv 
C^{\rm B1}_1(Q^2,r)|_{\mathcal{O}(\kappa^{-1})}\,$,
and
$\llbracket C^{\rm B1}_2(Q^2,r) \rrbracket_1 \equiv 
C^{\rm B1}_2(Q^2,r)|_{\mathcal{O}(\kappa^{-1})}\,$,
while $\llbracket C^{\rm B1}_1(Q^2,r) \rrbracket_1 = 
\llbracket C^{\rm B1}_2(Q^2,r) \rrbracket_0 = 0$.
To implement the rearrangement of endpoint-singular terms 
we introduce the scaleless integral
\begin{eqnarray}
\frac{2 C_F}{Q} \,f(\epsilon)\,|C^{\rm A0}|^2 \widetilde{\mathcal{J}}^{(\bar q)}_{\bar{c}}  
\widetilde{\mathcal{J}}_{c}^{(g)}  
\int_0^\infty d\omega d\omega'\,
\frac{D^{\rm B1}(\omega Q)}{\omega}
\frac{D^{\rm B1*}(\omega'Q)}{\omega'} 
\left \llbracket \widetilde{S}_{\rm NLP}(s_R,s_L,\omega,\omega')\right 
\rrbracket,
\label{eq:scaleless_integral}
\end{eqnarray}
which vanishes in $d$-dimensions. We split this 
integral in two terms $I_{1,2}$, $I_1+I_2=0$, with $I_1$ being defined 
by $\omega$ {\em or} $\omega'$ smaller than a parameter $\Lambda$ 
and $I_2$ the complement region, as shown on the left-hand side
of figure~\ref{fig:overlap}. 
In the complement region $\omega,\omega^\prime > \Lambda$, and the 
double-bracket asymptotic behaviour can be used for  
functions of $\omega,\omega'$ in the A-type term.
The endpoint rearrangement 
consists of subtracting $I_1$ from the B-type term 
and $I_2$ from the A-type term. The subtracted 
expressions are now separately endpoint-finite, but 
depend on $\Lambda$ which cancels exactly between the 
two terms as long as we do not make further approximations.
Starting with the A-type contribution,subtracting from it the complement region $I_2$ of the integral~\eqref{eq:scaleless_integral}, 
and using the endpoint factorization conditions results in 
\begin{eqnarray}
\frac{1}{\sigma_0}\frac{\widetilde{d\sigma}}{ds_R ds_L}|_{\rm A-type}
\! &=& 
\frac{2C_F}{Q} |C^{\rm A0}|^2 \,
\widetilde{\mathcal{J}}_{\bar c}^{(\bar q)} 
\,\int_0^\infty d\omega d\omega'\,
\bigg\{\,\widetilde{\mathcal{J}}_{c}(s_{L},\omega,\omega')
\,\widetilde{S}_{\rm NLP}(s_R,s_L,\omega,\omega')
\nonumber\\[0.2cm]\nonumber
&&\hspace*{-3.7cm}
-\,\theta(\omega-\Lambda)\theta(\omega'-\Lambda)\,
\llbracket\widetilde{\mathcal{J}}_{c}(s_{L},\omega,\omega')
 \rrbracket
\llbracket \widetilde{S}_{\rm NLP}(s_R,s_L,\omega,\omega')
\rrbracket 
\\[0.2cm]
&&\hspace*{-3.7cm}
+\;\,\widetilde{\!\!\widehat{\mathcal{J}}}_{\!c}(s_{L},
\omega,\omega') 
\,\,\widetilde{\!\widehat{S}}_{\rm NLP}(s_R,s_L,\omega,\omega')\,\bigg\}\,, \quad
\label{eq:Atype_subtracted_general2}
\end{eqnarray}
where we set $f(\epsilon)=1$ as $\epsilon\to 0$.
The remaining part $I_1$ of the integral \eqref{eq:scaleless_integral} is now combined
with  $i=i'=1$ part of the B-type term, and after using refactorization conditions we have
\begin{eqnarray}
\frac{1}{\sigma_0}\frac{\widetilde{d\sigma}}{ds_R ds_L}|_{
\scriptsize{ \begin{array}{l}
$\rm B--type$\\[-0.1cm] $i=i'=1$\end{array}}} 
\!&=& 
\frac{2C_F}{Q^2} \widetilde{\mathcal{J}}_{c}^{(g)} \,
\widetilde{S}^{(g)}(s_R,s_L)\, \int_0^\infty dr dr' 
\,\bigg[
 \nonumber\\ 
\nonumber
 &&\hspace*{-3.8cm}
\theta(1-r)\theta(1-r')
C^{\rm B1*}_{1}( r')C^{\rm B1}_{1}( r) 
\,\widetilde{\mathcal{J}}_{\bar{c}}^{q\bar{q}(8)}(s_R,r,r^\prime)\\[0.1cm]
 &&\hspace*{-3.8cm}
-
\big[1-\theta(r-\Lambda/Q)\theta(r'-\Lambda/Q)\big]\,
\llbracket C^{\rm B1*}_{1} ( r')\rrbracket_{0}\, 
\llbracket C^{\rm B1}_{1}( r) \rrbracket_{0} \, \llbracket \widetilde{\mathcal{J}}_{\bar{c}}^{q\bar{q}(8)}(s_R,r,r^\prime)\rrbracket_0
\bigg]\,.
\label{eq:Btype_subtracted2}
\end{eqnarray}
The $i=i'=2$ term with the anti-quark becoming soft takes a similar form and 
is explicitly provided in \cite{Beneke:2022obx}.
 
\section{Resummation}
\noindent Solving of the necessary RG equations for the objects appearing in the 
above factorization formulas is discussed in detail in sections 5.1 and
5.2 of \cite{Beneke:2022obx}. The final result 
for the leading-logarithmic accurate resummed expression after all the individual pieces combined together reads in Laplace space:
\begin{eqnarray}
	\frac{1}{\sigma_0}\frac{\widetilde{ d\sigma}}{ ds_R ds_L}|_{\rm LL}&=&
	2\cdot \frac{2 C_F}{Q s_R}\frac{\alpha_s(\mu_c)}{4\pi}\,
	\,\exp\left[4C_FS\klammer{\mu_h,\mu_{\bar{c}}}
	+4C_AS\klammer{\mu_s,\mu_c}\right]
	\times\klammer{\frac{Q^2}{\mu_h^2}}^{\!-2C_F A\klammer{\mu_h,\mu_{\bar{c}}}}
	\nonumber\\
	&&\hspace*{-1cm}
	\times	\klammer{\frac{1}{s_Ls_R e^{2\gamma_E}\mu_s^2}}^{\!-2C_A A\klammer{\mu_s,\mu_c}}
	 \int_\sigma^Q\frac{d\omega}{\omega}\,
	\klammer{\frac{\omega }{s_R e^{\gamma_E} \mu_{s\Lambda}^2}}^{\!-2\klammer{C_F-C_A}A\klammer{\mu_{s\Lambda},\mu_{h\Lambda}}}
	\nonumber\\[0.2cm]
	&&\hspace*{-0.8cm}\times
		\exp\left[4\klammer{C_F-C_A}S\klammer{\mu_{s\Lambda},\mu_{h\Lambda}}\right]
	\,\klammer{s_R e^{\gamma_E} Q}^{2C_FA\klammer{\mu_{h\Lambda},\mu_{\bar{c}}}+2C_A A\klammer{\mu_c,\mu_{h\Lambda}}}
	\,,
	\label{eq:mainLLresult}
\end{eqnarray}
with the functions $S\klammer{\nu,\mu}$ and $A_{\gamma_i}\klammer{\nu,\mu}$ defined as in \cite{Neubert:2004dd}. We note that $\mu^2_{h\Lambda}\sim \omega Q$ and $\mu^2_{s\Lambda}\sim \omega/s_R$ are $\omega$-dependent scales 
that appear inside the integrand for reasons explained in \cite{Beneke:2022obx}. The importance of  
{\em next}-to-leading logarithms can be studied by varying the
various matching scales around the 
values adopted in \eqref{eq:mainLLresult}. We 
vary the three pairs of scales $(\mu_h, \mu_{h\Lambda})$, 
$(\mu_c,\mu_{\bar c})$, $(\mu_s, \mu_{s\Lambda})$ by a factor of $1/2$ 
and 2 around their default scales. Taking the minimum 
and maximum values to compute the scale variation. 
We show this for the normalized Laplace-space distribution 
$\frac{Q s}{\sigma_0}\frac{\widetilde{ d\sigma}}{ ds}$ in the 
right-hand panel of figure~\ref{fig:overlap} as the light-red 
band around the red curve (LL) that represents 
\eqref{eq:mainLLresult}. For comparison, the tree-level (LO) and 
linear-$\beta_0$ truncation
of the LL expression are displayed. 
The sizeable scale variation in the figure emphasizes the need for 
NLL resummation. The endpoint-rearranged 
factorization formula presented in this work provides the 
starting point for this systematic improvement.

\section{Conclusion}

\noindent In this work, we summarized the derivation of a novel endpoint factorization relation for 
the  NLP gluon-thrust distribution in the two-jet region, as achieved in \cite{Beneke:2022obx}. 
This off-diagonal contribution contains a gluon-initiated jet recoiling against 
a quark-antiquark pair, which involves subleading-power cross-section level soft and jet 
functions.
The framework shares many similarities with the rearrangement employed for the resummation of the $H\to \gamma\gamma$ bottom-loop amplitude  
\cite{Liu:2019oav,Liu:2020wbn} and 
allows for the first time to systematically remove 
endpoint divergences in the convolution integrals of SCET$_{\rm I}$
factorization theorems, opening the path 
to NLP resummation for collider observables with soft quark emission.
Employing the subtraction of endpoint divergences with the aid of operatorial
factorization conditions, we managed to reshuffle the factorization 
theorem such that the individual terms are free from 
endpoint divergences and can be written in 
terms of renormalized hard, (anti) collinear, and soft functions in four 
dimensions. At this point, standard RG 
techniques can be applied to obtain the resummed integrands.
In \cite{Beneke:2022obx}, we derived the anomalous dimensions of the NLP jet 
and soft functions using RG consistency and endpoint 
factorization relations. We also calculated the one-loop anomalous dimension for the hard matching coefficients, which enabled us to perform the first resummation of the endpoint-divergent 
SCET$_{\rm I}$ observable at the LL accuracy using exclusively RG methods.  We verified that our results can recover the earlier results at DL accuracy \cite{Moult:2019uhz,Beneke:2020ibj}
and we evaluated the numerical impact of the LL corrections. Our main result for gluon-thrust is 
\eqref{eq:mainLLresult}, which provides an expression for 
the two-hemisphere invariant mass distribution in Laplace space.
The next step in further exploration of NLP resummation 
is the extension of the present work and related 
soft-quark emission processes to NLL accuracy,
which requires the calculation of renormalization kernels for the NLP soft and jet functions at order $\mathcal{O}(\alpha_s)$, and application of the methods presented here to
address endpoint divergent issues appearing in diagonal channels of relevant processes.

\subsubsection*{Acknowledgements} 
\noindent This work has been supported by the Excellence Cluster 
ORIGINS funded by the Deutsche Forschungsgemeinschaft 
(DFG, German Research Foundation) under Ger\-many's Excellence Strategy --EXC-2094 --390783311.
S.J. is supported by the UK Science and Technology Facilities Council under grant ST/T001011/1.
R.S. is supported by the United States Department of Energy under Grant Contract DE-SC0012704.
L.V. is funded by Fellini Fellowship for Innovation at INFN, and by the European Union's Horizon
2020 research programme under the Marie Sk\l{}odowska-Curie Cofund Action, grant agreement no.
754496. J.W. is funded by the National Natural Science Foundation of China (No.~12005117) 
and the Taishan Scholar Foundation of Shandong province (tsqn201909011).

\bibliographystyle{JHEP}
\bibliography{NLPthrustUpdate220708}

\end{document}